\documentclass[conference,comsoc,9pt,a4paper]{IEEEtran}
\usepackage{amsmath,amssymb,amsfonts}
\usepackage{algorithmic}
\usepackage{graphicx}
\usepackage{textcomp}
\usepackage{xcolor}
\usepackage{hyperref}
\usepackage{listings, listings-rust}
\usepackage{listing}
\usepackage[sorting=none]{biblatex}
\def\BibTeX{{\rm B\kern-.05em{\sc i\kern-.025em b}\kern-.08em
    T\kern-.1667em\lower.7ex\hbox{E}\kern-.125emX}}
\usepackage{listings, xcolor}

\definecolor{verylightgray}{rgb}{.97,.97,.97}

\lstdefinelanguage{Solidity}{
	keywords=[1]{anonymous, assembly, assert, balance, break, call, callcode, case, catch, class, constant, continue, constructor, contract, debugger, default, delegatecall, delete, do, else, emit, event, experimental, export, external, false, finally, for, function, gas, if, implements, import, in, indexed, instanceof, interface, internal, is, length, library, log0, log1, log2, log3, log4, memory, modifier, new, payable, pragma, private, protected, public, pure, push, require, return, returns, revert, selfdestruct, send, solidity, storage, struct, suicide, super, switch, then, this, throw, transfer, true, try, typeof, using, value, view, while, with, addmod, ecrecover, keccak256, mulmod, ripemd160, sha256, sha3}, % generic keywords including crypto operations
	keywordstyle=[1]\color{blue}\bfseries,
	keywords=[2]{address, bool, byte, bytes, bytes1, bytes2, bytes3, bytes4, bytes5, bytes6, bytes7, bytes8, bytes9, bytes10, bytes11, bytes12, bytes13, bytes14, bytes15, bytes16, bytes17, bytes18, bytes19, bytes20, bytes21, bytes22, bytes23, bytes24, bytes25, bytes26, bytes27, bytes28, bytes29, bytes30, bytes31, bytes32, enum, int, int8, int16, int24, int32, int40, int48, int56, int64, int72, int80, int88, int96, int104, int112, int120, int128, int136, int144, int152, int160, int168, int176, int184, int192, int200, int208, int216, int224, int232, int240, int248, int256, mapping, string, uint, uint8, uint16, uint24, uint32, uint40, uint48, uint56, uint64, uint72, uint80, uint88, uint96, uint104, uint112, uint120, uint128, uint136, uint144, uint152, uint160, uint168, uint176, uint184, uint192, uint200, uint208, uint216, uint224, uint232, uint240, uint248, uint256, var, void, ether, finney, szabo, wei, days, hours, minutes, seconds, weeks, years},	% types; money and time units
	keywordstyle=[2]\color{teal}\bfseries,
	keywords=[3]{block, blockhash, coinbase, difficulty, gaslimit, number, timestamp, msg, data, gas, sender, sig, value, now, tx, gasprice, origin},	% environment variables
	keywordstyle=[3]\color{violet}\bfseries,
	identifierstyle=\color{black},
	sensitive=true,
	comment=[l]{//},
	morecomment=[s]{/*}{*/},
	commentstyle=\color{gray}\ttfamily,
	stringstyle=\color{red}\ttfamily,
	morestring=[b]',
	morestring=[b]"
}

\begin{document}

\title{Towards Proxy Staking Accounts Based on NFTs in Ethereum}

\ifdefined\DOUBLEBLIND
  \author{\IEEEauthorblockN{Author Name}
\IEEEauthorblockA{\textit{Affilliation} \\
\textit{Affiliation}\\ 
City and Country \\
email}
}
\else
\author{\IEEEauthorblockN{Viktor Valaštín, Roman Bitarovský, Kristián Košťál and Ivan Kotuliak}
\IEEEauthorblockA{\textit{Faculty of Informatics and Information Technologies} \\
\textit{Slovak University of Technology in Bratislava}\\
Bratislava, Slovakia \\
\{viktor.valastin, xbitarovsky, kristian.kostal, ivan.kotuliak\}@stuba.sk}
}
\fi

\maketitle

\begin{abstract}
Blockchain is a technology that is often used to share data and assets. However, in the decentralized ecosystem, blockchain-based systems can be utilized to share information and assets without the traditional barriers associated with solo responsibility, e.g., multi-sig wallets. This paper describes an innovative approach to blockchain networks based on a non-fungible token that behaves as an account (NFTAA). The key novelty of this article is using NFTAA to leverage the unique properties of NFTs to manage your ownership better and effectively isolate them to improve the security, transparency, and even interoperability possibilities. Additionally, the account-based solution gives us the ability and flexibility to cover regular use cases such as staking and liquid equities, but also practical composability. This article offers a simple implementation, which allows developers and researchers to choose the best solution for their needs in demand of abstract representation in any use case.
\end{abstract}


\section{Introduction}
When Ethereum transitioned to the Proof-of-Stake consensus, it was a huge signal for the ecosystem \cite{Kapengut2022}. This switch from the Proof-of-Work (PoW) consensus mechanism is being done for several reasons. Firstly, PoS is more energy efficient and environmentally friendly than PoW \cite{Schinckus2022}. It also allows for faster transaction processing and improved scalability. PoS allows for more decentralized decision-making, as the power to validate transactions is distributed among multiple "mining" nodes rather than concentrated in a few large mining pools \cite{Johnson2014}.

To ensure correct protocol mechanisms, asset staking was invented. 
Staking refers to holding and committing assets as collateral to participate in the validation of transactions and the production of new blocks. Stakers, also known as validators, are producing new blocks for blockchain. If a validator produces a valid block, they are rewarded with a portion of the transaction fees and a block reward. However, if they produce an invalid block, they may be penalized, and their staked assets may be slashed. 

Yet, staking has a few problems, and in addition to discussing and speculating about its decentralization or securing, one of them is the illiquidity of staked assets and the impossibility of changing their owner. The inability to change the owner of staked assets can be very inconvenient, for example, if we urgently need to exchange our current staked collateral for other assets.
This paper closely looks at this problem, analyzes the existing solutions, and describes a new concept that introduces the staking assets through an NFT account (NFTAA), i.e., transferring the ownership of the stake to the NFT account, which allows changing the owner of the stake (through a change of ownership of the NFTAA) in case that we need.

The rest of the paper is organized as follows:
In Section \ref{sec:ethintro}, a closer look at the example of Ethereum and its technical specifications is provided. Existing related works are analyzed in Section \ref{sec:related}. Next, the solution is presented in the form of NFT proxy accounts described in Section \ref{sec:design}, and our evaluation is described in Section \ref{sec:evaluation}. We summarize the paper, the discussion, and future work in Section \ref{sec:summary}.

\section{Ethereum introduction}\label{sec:ethintro}
Ethereum stands tall as the unrivaled titan in the realm of smart contract platforms, commanding its position as the foremost and most sophisticated hub for decentralized financial ecosystems across the globe.

This pioneering blockchain network is purpose-built to facilitate the seamless operation of decentralized applications (dApps) while steadfastly pursuing continuous enhancements in scalability, an ongoing mission undertaken by its vibrant community of developers and stakeholders. Ethereum's commitment to scalability underscores its unwavering dedication to delivering a future-proof infrastructure that can efficiently accommodate the ever-expanding demands of the decentralized world. Through its tireless innovation, Ethereum endeavors to remain the largest, most adaptable, and responsive smart contract platform, serving as the bedrock of the decentralized digital economy.

%\subsection{Network node \label{subsec:node}}
%Running your own node is always a good idea when interacting with blockchain networks, and we can provide better privacy, security, resilience to potential censorship, support the network, and increase decentralization, but it is unnecessary if we want to interact with Ethereum.

%Unlike conventional nodes, a significant amount of technical skills, such as working with the command line, are required to operate them. Ethereum provides DAppNode, free and open-source software that gives users an app-like experience while managing our node \cite{runanode}.

%Although running our node is not necessary, we will not avoid it if we want to apply our ETH in staking and thus contribute to network security and participate in the validation and creation of new blocks.

\subsection{Proof of stake \label{subsec:pos}}
In proof of stake (PoS), compared to proof of work (PoW) used in Bitcoin, we do not have miners but rather validators.
Validators are nodes that validate blocks and their data, such as transactions. 
Validators are not competing to find the correct hash for a block and do not receive a reward in the form of a coinbase transaction as miners in Bitcoin.
Instead, validators receive a reward through transaction fees from the transactions included in the block they validate.
Having a lot of computing power to be a validator is unnecessary, and the process is less energy-intensive \cite{blockchainWithoutWastePoS}.

Validators are motivated to refrain from cheating because they must stake a certain amount of coins to become validators. 
These staked coins are used to secure the network and safeguard against cheating. 
If a validator is caught cheating, they lose their staked coins and can no longer serve as a validator. 
Since the reward for validating a block is typically less than the amount of staked coins, it is not financially worthwhile for a validator to cheat. 
In Ethereum, the amount of staked coins required to become a validator is set at 32 ETH \cite{ethereumTransitionToPoS}. Staking this amount is quick; however, unstaking is slightly different and can take days; recently, it was more than five days \cite{nijkerk_ethereum_2024}. Therefore, selling the account with staked 32 ETH is quicker and gives users easy and fast access to money. However, you do not want to sell your original account because you could have history and other liquidity associated with it. Thus, we are introducing staking through an NFT which works as a proxy and isolates your stake, so you can easily sell it to anyone without risk of selling more than your stake.

% \subsection{Smart contracts \label{subsec:smartcontracttheory}} 
% Smart contracts are simply pieces of code stored on the blockchain and executed by it. 
% Once a smart contract is deployed, its code cannot be changed. 
% In many cases, smart contracts are self-executing and do not require human interaction.

% However, smart contracts provide services that can replace traditional paper contracts. 
% In smart contracts, the parties involved are not individuals but addresses. 
% Since everything on the blockchain is immutable, once a smart contract is deployed and executed, it cannot be changed, and 
% everything it defines is considered the truth. Additionally, smart contracts are transparent, as everything on the Ethereum blockchain is public \cite{masteringEthereum}.

% There are potential issues to be aware of when using smart contracts. For example, if a poorly written smart contract can be exploited, and 
% of the funds it holds can be stolen. Once this occurs, the stolen funds are lost permanently, as smart contracts and all transactions on the blockchain are immutable.
% Moreover, smart contracts can be used for more than just transferring funds. They can also be utilized for tasks such as voting and insurance, among many others \cite{khan2021blockchain}.

\section{Related work and Background} \label{sec:related}
As mentioned, proof-of-stake can provide security for the network by other participants.
As a bonus, there is also gaining of some profit from the staking. However, PoS has some drawbacks.
First, to unstake the tokens, there is a need to wait for the withdrawal period.

The current withdrawal mechanism for validators has shown significant scalability concerns as the network grows. At present, the system only accommodates 16 withdrawals within a single block. This translates to a maximum processing capability of 115,200 validator withdrawals daily under the ideal scenario of no missed slots. This limitation becomes particularly pronounced in light of rising validator numbers: 800,000 withdrawals take an entire week to process. This rigid structure poses a severe challenge as it correlates the number of validators with the withdrawal processing time.

Moreover, the inherent risk of missed slots further complicates the landscape. While the system is designed to bypass validators without eligible withdrawals, thereby potentially saving some time, the downside of missed slots has a pronounced effect. Each missed slot can lead to a proportional delay in processing, pushing the timeline even further. Such inefficiencies can negatively impact user experience, potentially dissuading prospective validators from participating and undermining the very foundation of the network's decentralization principle. This bottleneck, if not addressed, threatens both the operational efficacy and the perceived reliability of the platform.

Fortunately, already existing solutions can help solve these problems.
%This section describes the most popular solutions currently used in the Ethereum ecosystem.
Generally speaking, there is a familiar concept: Lock ETHs and get a token representing the staked amount (derivate). Getting derivate from the staked asset is also called liquid staking. It represents the idea of making locked assets liquid. A liquid asset represents an asset that can be converted to cash. In DeFi terms, liquid assets can be swapped for other assets, such as stablecoins, or provide additional liquidity on the various decentralized exchanges.

\subsection{Liquid Staking Protocols}
Liquid staking providers empower you to tap into the liquidity of your staked assets. In return, these providers typically claim a share of the staking rewards ranging from 5\% to 10\% as compensation for rendering this valuable service.
With the Shanghai upgrade, the appeal of Ethereum Liquid Staking Derivatives (LSDs) is set to soar, as it alleviates the inherent risks and inconveniences associated with conventional ETH staking. This surge in LSD adoption translates into amplified revenue streams for liquid staking platforms, instilling a bullish sentiment across the liquid staking sector.
Anticipating this upward trajectory, MetaMask has thoughtfully integrated staking directly into its wallet interface, streamlining the process for users to stake their ETH. Users can look forward to earning approximately 5\% annual rewards in ETH by leveraging this feature, with support currently extended to Lido and Rocket Pool and the possibility of more integrations in the pipeline \cite{lsdcoingeco}.

\subsection{Lido}
Lido made its mark as a trailblazer in the field of liquid staking. In 2020, with the launch of Ethereum's Beacon chain, Lido became the first provider to offer liquid staking services, pioneering and popularizing this innovative approach. Since its inception, Lido has consistently maintained its position as a market leader.

Lido's liquid staking solution revolves around its token known as stETH, which represents users' staked Ethereum (ETH). The process is simple: users deposit their ETH on Lido's staking platform and receive stETH in return. This transformative token empowers users by preserving the liquidity of their staked assets. With stETH in hand, users can explore a myriad of DeFi opportunities.
Holders of stETH can choose to engage in various DeFi activities, including trading, lending, providing liquidity, and using it as collateral. Additionally, stETH holders earn ETH staking rewards in the form of more stETH tokens, further enhancing their potential yield.

Lido operates on a sustainable revenue model. The platform collects a 10\% fee from the ETH staking rewards generated. This fee is divided between Lido's node operators, who ensure the network's security, and the DAO (Decentralized Autonomous Organization) treasury.

Despite its outstanding achievements, Lido's dominant position has raised concerns within the Ethereum community. Concentrating a substantial amount of Ether within a single organization poses potential risks to the network. A centralized attack on the Ethereum network could become feasible if Lido's control surpasses 33\%. This scenario, known as a 1/3 centralization attack, is a recognized challenge in Proof-of-Stake ecosystems. In response, Lido and the Ethereum community are actively exploring avenues to enhance the decentralization of the protocol, mitigating these potential risks \cite{lsdcoingeco, lidodocx}.

\subsection{Rocket Pool}
Rocket Pool is a groundbreaking Ethereum PoS protocol designed to be community-owned, decentralized, trustless, and compatible with Ethereum's staking system. The project was conceived in late 2016 and has been live since October 2021. Rocket Pool caters to two primary user groups:

\begin{itemize}
    \item Tokenized Stakers: Individuals who want to participate in Ethereum staking using rETH with deposits as low as 0.01 ETH.
    \item Node Operators: Users who stake ETH and run a node in the Rocket Pool network to earn higher returns compared to staking independently outside the protocol. This includes experienced users and those who may not have 32 ETH for solo staking.
\end{itemize}

The core premise of Rocket Pool is to ensure the network remains decentralized, aligning with Ethereum's ethos and promoting self-sovereignty.
Rocket Pool comprises several key components that create a decentralized Ethereum staking ecosystem. \textbf{Smart contracts} facilitate ETH deposits, allocate them to node operators based on demand, and manage various tokens and rewards. Rocket Pool's decentralized network of specialized Ethereum nodes runs the \textbf{Smart Node} software. These nodes communicate with the protocol's smart contracts and provide the network consensus required by the Ethereum Beacon Chain. \textbf{Minipool validators} are smart contracts created by node operators who deposit 8 or 16 ETH on their node. When the total ETH in a Minipool validator reaches 32 ETH, a new validator is created to perform consensus duties.

Rocket Pool's commitment to decentralization, security, and inclusivity makes it an attractive solution for those looking to participate in Ethereum staking without the traditional barriers associated with solo staking. The decentralized protocol ensures resilience, scalability, and more equitable distribution of rewards within the network \cite{lsdcoingeco, rocketpool}.

\subsection{EIP-6551} \label{subsec:6551}
The world of non-fungible tokens (NFTs) is on the brink of a significant transformation with the introduction of the ERC-6551 token standard. This new standard seeks to enhance the functionality of NFTs by equipping them with smart contract capabilities, allowing them to represent unique digital assets and act as smart contract wallets.
The ERC-721 standard, the foundation for creating unique digital assets on the Ethereum blockchain, has its limitations. While it paved the way for myriad applications, from art to virtual real estate, its scope is restricted. NFTs based on ERC-721 can only be owned and transferred; they cannot own other assets or interact dynamically with other smart contracts.
Proposed as an Ethereum Improvement Proposal (EIP) in February 2023, ERC-6551 aims to address these limitations. It introduces a system where every NFT can be assigned an Ethereum account, termed a token-bound account (TBA). This account enables the NFT to own assets, interact with applications, and execute arbitrary operations, all without necessitating changes to existing smart contracts or infrastructure. Key features:

\begin{itemize}
    \item Smart Contract Capabilities: Beyond representing unique assets, ERC-6551 tokens can function as smart contract wallets, holding other tokens and NFTs.
    \item Interactivity: TBAs can seamlessly interact with various entities on the Ethereum network, from decentralized exchanges to gaming platforms.
    \item Composability: This feature allows NFTs to bundle with related assets, enhancing user experiences, especially in Web3 gaming environments.
    \item Identity \& Provenance: Each NFT, with its distinct identity, can interact independently with decentralized applications (dApps). Additionally, users can trace the complete transaction history of an asset, offering a clearer picture of its past interactions and uses.
\end{itemize}

The ERC-6551 standard represents a pivotal moment in the evolution of NFTs. Bridging the gap between unique digital assets and smart contract functionality opens up a realm of possibilities in the Web3 space, from gaming to art and beyond. As the crypto community navigates its challenges, the potential of ERC-6551 to redefine digital ownership remains immense \cite{eip6551, tokengame, eipexpt}.

But this solution also has some gaps. Firstly, the existence of TBA is not documented in the NFT, meaning we could possess an NFT without any knowledge of an associated account until we check the TBA register. We can also sell an NFT with possession of different assets by mistake and thus lose our precious liquidity. The second major gap is the lack of atomicity in the creation process. Of course, this is because TBA applies to all existing NFTs. Still, we think that creating an NFT and wallet in one atomic transaction, ensuring that the NFT is fully bound to the wallet proxy contract from its inception and that the NFT has information about its connection to the wallet contract, is more transparent.

\section{Solution design} \label{sec:design}
In the case of the existing solutions (except ERC-6551), the ETH collateral is often protected via a DAO / multi-signature account.
However, this means that the security of the assets is not directly in the hands of the user because, in case of a problem or a third-party devaluation, our liquid assets representing our stake are depreciated.
This section delves into the cutting-edge notion of staking via NFTs, which serve as distinct accounts within the system. Although this method is not yet common, we bring a few insights into why it is a good idea. As we see it as a non-standard idea, the novelty of our work lies in exploring its potential, while considering this rarity not as a weakness, but as an opportunity to innovate in the field of blockchain applications.

The term "non-fungible token" refers to a unique and distinct asset that can represent various physical items with varying values. On the other hand, NFTs primarily operate on the asset side of the owner-asset relationship and lack the potential to own other assets. On the other hand, accounts are used to represent ownership of assets. In an owner-asset relationship, the account can be on the first or second side.
After composing these two relations, it follows that it is possible for an account and an NFT to be represented by a single object that will be uniquely identifiable and will have the nature of an asset that can be owned - change its owner and at the same time can own other assets.
We represent a non-fungible token as an account, NFTAA in short.
Most Ethereum is based on a single account architecture consistent with the NFTAA philosophy and facilitates connection with existing infrastructure. We may easily exploit existing infrastructure for smooth communication if NFTAAs are handled as conventional accounts. Furthermore, if needed, NFTAAs can function as asset owners or in a multi-sig scheme.

\begin{figure}[htbp]
    \centering
    \includegraphics[width=0.2\textwidth]{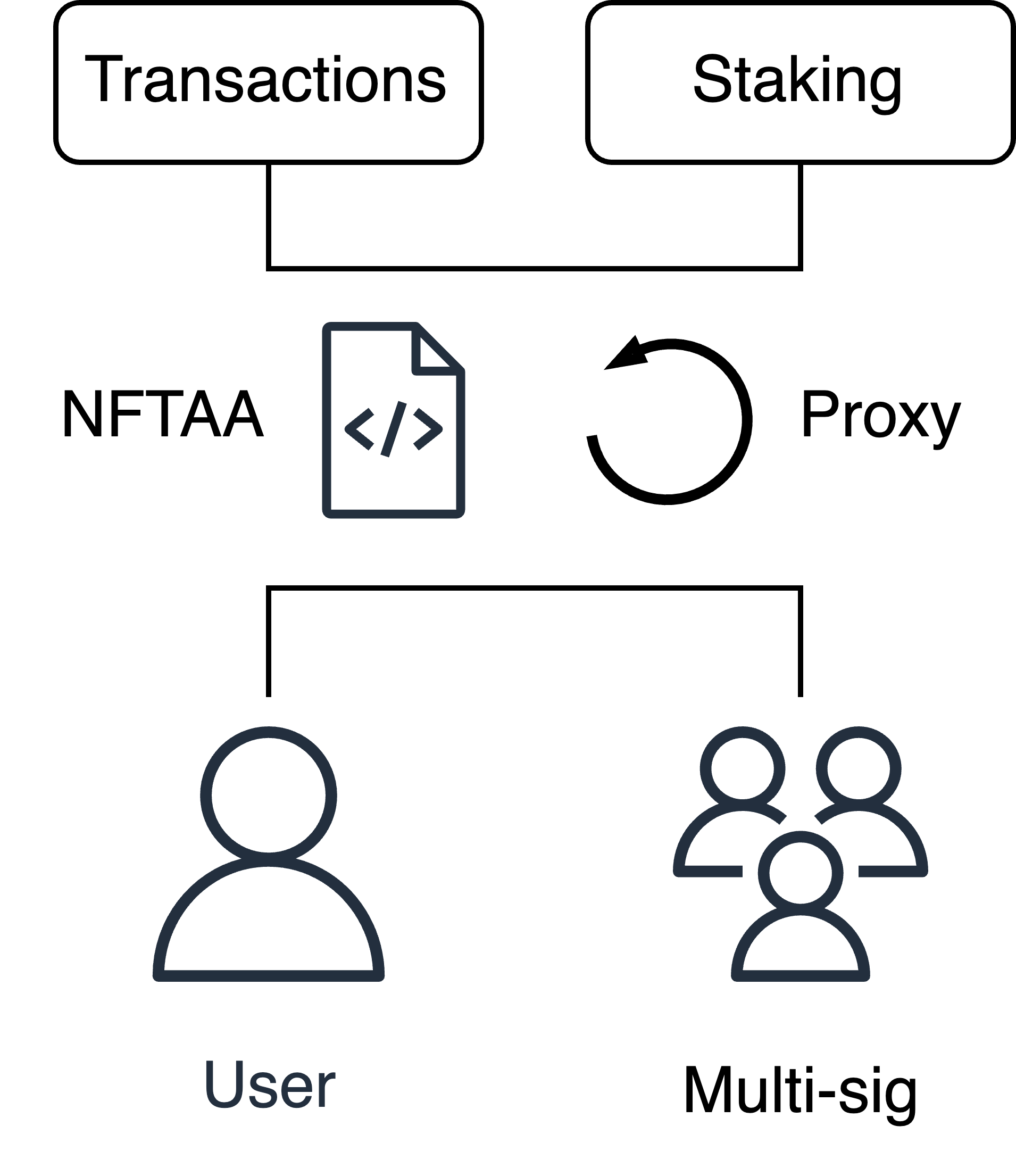}
    \caption{NFT as an account interaction}
    \label{fig:nftaa-sc-proxy}
\end{figure}

Another key characteristic of NFTAA is the ownership possibilities and the transferability of this ownership to any other account, potentially other NFTAAs. This means that the private keys used to alter a certain NFTAA will change.
This is feasible thanks to the well-known proxy smart contracts \cite{EthProxy, OpenProx}. Figure \ref{fig:nftaa-sc-proxy} shows the relationships between NFTAA and other logical entities.

We can create a proxy account (smart contract) to perform specific tasks on behalf of another account. For example, Bob's account can have a proxy account named Alice. Alice can send transactions on Bob's behalf, but only for the calls Bob has approved, as Bob is the only person currently authorized to access the proxy. This means that if Alice sends a transaction, it is initiated by Bob, so Bob appears to be signed by Bob, even though Alice is technically sending it.

\begin{figure*}[htbp]
    \centering
    \includegraphics[width=0.7\linewidth]{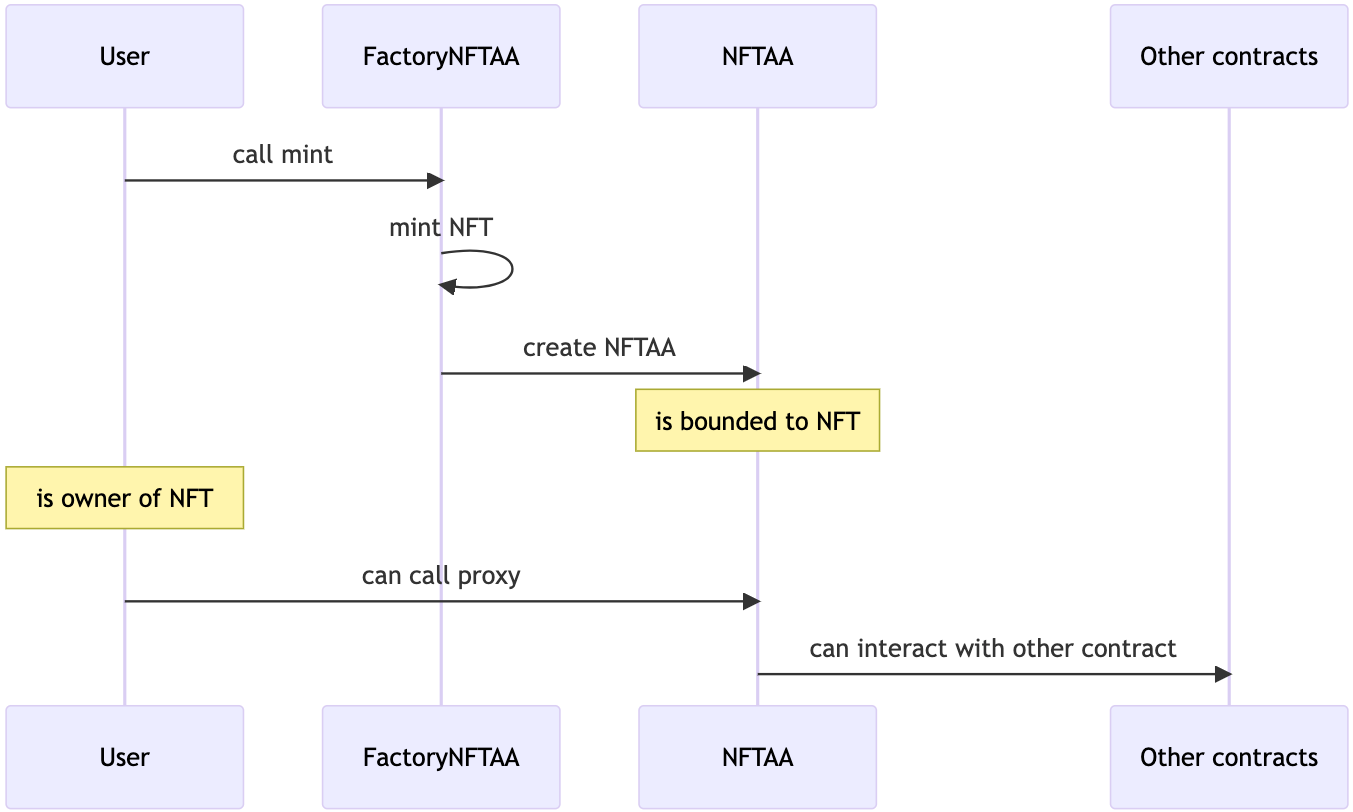}
    \caption{NFTAA Solidity sequence diagram}
    \label{fig:nftaa-sol-sq}
\end{figure*}

\subsection{DeFi and DAO use cases \label{sec:defidaousecases}}
When we discuss accounts, NFTs, assets, staking, ownership, and owners - DeFi comes into play, and because of the nature of NFTAA, we will also discuss DAO.
As stated before, if we own ETH and decide on liquidity staking, we have staked the ETH in exchange for other assets we can continue utilizing.
Compared to futures or tokens as a reward for staking, NFT may appear to be one large illiquid piece of value. This is not the case because fractionalizing NFT is not novel in the DeFi world \cite{fractional}.
The ideal number of fractions equals the number of ETH tokens staked on behalf of the NFTAA account. However, for whatever reason, we may be unable to fold NFTs distributed in this manner back to their original condition.
We could remedy this by holding a buyout auction or depositing a sum greater than the reported staking value. 
%The use-cases for NFTAA are shown in Figure \ref{fig:usecasedia}.
%\begin{figure}[htbp]
%    \centering
%    \includegraphics[width=\columnwidth]{images/usecasediagram.png}
%    \caption{NFTAA use-case diagram}
%    \label{fig:usecasedia}
%\end{figure}
We can still perform an unstake if our NFTAA becomes fully illiquid for any reason and cannot be sold on the secondary market or used as collateral for a loan, and after this, the NFTAA may (or may not) be burnt.

Another intriguing application of NFTAA is the ability to manage an account using numerous proxy accounts, each with the capacity to make certain calls. This offers numerous benefits as we see it as a new paradigm for asset management and ownership in the ecosystem of decentralized finances. 
We can think of NFTAA as a company with employees who have specific roles. Specific subaccounts represent particular roles, but the company signs all transactions. From another point of view, NFTAA represents a company that can hold shares in the form of investments and other liquid or unique assets. 
NFTAA, like any other corporation, can be acquired. The NFTAA also functions as the company's ledger. 

Last but not least, the big trend in the Ethereum ecosystem is re-staking liquid staking tokens (LRST), such as Lido and Rocket Pool mentioned in Section \ref{sec:related} as re-staked tokens serve as a security for data availability layers - mainly dominated by EigenLayer \cite{eigenlayer}. We open new possibilities and scientific contributions of LRSTs because the NFTs can give better portfolio control without emitting new tokens for re-staking.

\subsection{Architecture}
Figure \ref{fig:nftaa-sol-sq} shows the key elements of the architecture. The foundation of the entire implementation is the NFTAA smart contract, which has ERC-721 as the underlying implementation. Subsequently, the NFT mining functionality is enriched by the creation of the NFTAA smart contract, which is interleaved from the NFT that has been mined. The NFTAA contract contains a proxy function that allows the owner of the pinned NFT to perform transactions through the NFTAA account. 

Listing \ref{lst:onlynft} shows how we can modify a function to be successfully called only if the caller owns a bound NFT.

\begin{lstlisting}[language=Solidity,caption={NFT owner only modifier},label={lst:onlynft}]
modifier onlyNFTOwner() {
    require(
        msg.sender == ERC721(_binded_nft_addr).ownerOf(_binded_nft_id),
        "Caller is not the owner of the NFT"
    );
    _;
}
\end{lstlisting}

\begin{figure*}[htbp]
    \centering
    \includegraphics[width=0.75\linewidth]{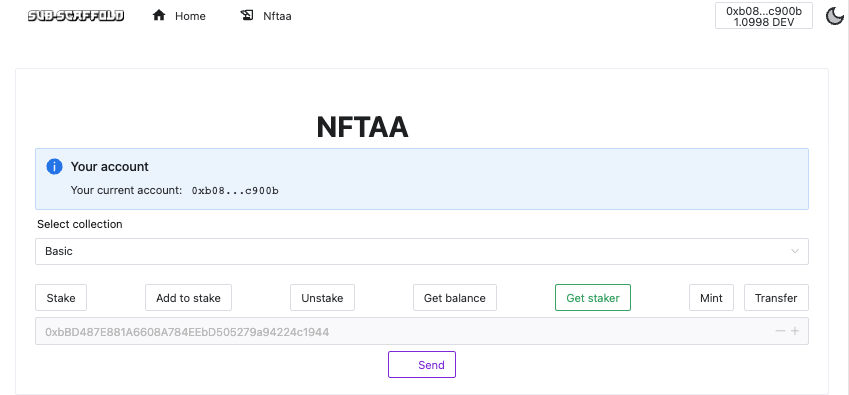}
    \caption{NFTAA dApp}
    \label{fig:nftaadapp}
\end{figure*}

We created a web application that allows users to engage with the NFTAA manufacturing contract. The application is built on a template for developing web apps that interface with web3 applications \cite{subscaffold}. This template is useful, but it only works with the Talisman wallet. Unfortunately, the Talisman documentation does not state if it allows signing EVM transactions \cite{talismanevm, talismanfaq}. To alleviate this constraint, we have added Metamask wallet support.
Finally, for ease of use, as we can see in Figure \ref{fig:nftaadapp}, we created a dApp using the NFTAA concept for staking. Via this decentralized application, we can:
Stake,
 Add other funds to the stake,
 Unstake,
 Get a balance of the stake,
 Get the staker address,
 Mint new NFTAA,
 Transfer ownership of NFTAA.

NFTAA introduces a novel approach to account management. Each user is issued a unique NFT that can represent their account. This NFT can encode specific attributes, permissions, or properties related to the user, such as staking rights, access to certain functionalities, or governance privileges within a decentralized ecosystem.
In particular, the use case of proxy-staking accounts leverages this NFT-based approach for staking mechanisms. Instead of the conventional method of staking tokens in a wallet, users stake their NFT-based accounts' tokens. The NFT encapsulates staking parameters, delegation preferences, and other relevant details for participating in a proof-of-stake (PoS) or delegated proof-of-stake (DPoS) consensus mechanism.
Through this model, the NFT serves as a representation of ownership and a means to determine the user's staking weight and influence within the network. The more tokens staked through a bound NFT account, the more significant the user's impact on consensus, rewards, and potentially, governance decisions.

Furthermore, NFTAA can facilitate delegation strategies, allowing users to delegate their staking rights or influence to other accounts or entities while still maintaining ownership of their NFT-based account. This flexibility enhances the overall decentralization and participation in the network.
This proposed framework does not emphasize the preservation of backward compatibility with existing NFT contracts, abstaining from alterations to the ERC-721 standard.
However, it is theoretically possible to choose the standard to be used when creating NFTAA so that we could use ERC-165, but of course, we prefer to use ERC-721. Alternatively, a partial subset of the ERC-721 interface could be used (e.g., ENS NameWrapper names).

\subsection{NFTAA vs TBA}
As highlighted in Section \ref{subsec:6551}, EIP 6551 presents certain limitations. Our solution, however, does not prioritize connecting existing NFTs but enables bidirectional binding of the NFT and Proxy contract. Consequently, upon ownership of an NFT, we can discern from the NFT metadata that it is associated with another smart contract, indicating its status as an NFTAA. Furthermore, we enhance transparency by ensuring the atomic creation of the NFT and proxy contract within a single transaction, thereby promoting seamless integration. Likewise, our design does not need the registrar service to create the TBA address as opposed to TBA. In our case, the factory contract creates the NFTAA contract with functionality that also works as the NFTAA address (TBA uses one more address than NFTAA). Another difference is that in TBA, you could have more accounts on one NFT, which in our case is not possible due to the atomic creation of exactly one NFT at the time of creating the NFTAA smart contract (we first create the account contract and exactly after that the unique NFT). However, theoretically, we could use already existing NFTAA as the NFT in TBA :-). 
Regarding security, there are at least two differences between NFTAA and TBA. The first is the possibility of generating the address of TBA sooner without having the deployed contract (as the create2 function for creating contracts is deterministic, i.e., you can calculate the future contract address). In this case, if the upcoming smart contract is badly developed (does not contain an execute function for byte calls to other contracts) or is not upgradeable, it can happen that assets or liquidity sent to the TBA will not be withdrawable and become locked forever (even the NFT bounded in TBA can be sent by a mistake (if not safeguarded) to the address of TBA. Therefore you lock yourself from it). The problem is that TBA uses only the address of any contract you want in the registry contract and does not look for any details on how the contract looks or if it is already deployed. In our case, the account form is pre-defined in the factory contract, and even after deployment, we guarantee upgradeability as it is a proxy account. 
The second security issue is potential fraud, which is described in the EIP 6551 specification \footnote{https://eips.ethereum.org/EIPS/eip-6551\#fraud-prevention} that you withdraw assets from the TBA and at the same time sell the NFT to someone who fought that they will also have the assets in the TBA. In our case, we provide the security measures for not withdrawing the assets from the NFTAA while also selling the NFT, which is bound to the NFTAA smart contract. A small difference is also in the address where you receive the assets (unstake in the case of staking). In TBA, the unstaked amount will go to the TBA address; in NFTAA, the unstaked amount will go to the NFTAA smart contract address.

\section{Evaluation and testing} \label{sec:evaluation}

When testing decentralized applications, the most important thing to focus on is the core functionality performed on the blockchain. 
The most significant part is the smart contract. Testing smart contracts is more complicated than testing regular software, but many tools make this easier nowadays. 
One of them is the Ethereum development environment Hardhat \cite{hardhat}. 
This development tool allows us to run the blockchain network locally and, among other things, provides extensive support for testing and mocking data to simulate a real blockchain environment. We provide the source codes on GitHub\footnote{https://github.com/fiit-ba/NFTAA}.

\subsection{Smart contract testing}
\begin{figure}[htbp]
    \centering
    \includegraphics[width=\columnwidth]{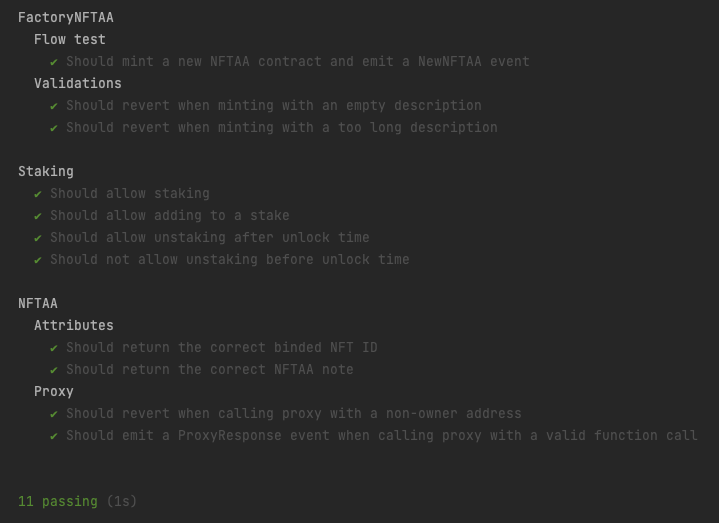}
    \caption{Results of smart contracts testing}
    \label{fig:tests}
\end{figure}

Based on these functionalities, we built our tests documented in Figure \ref{fig:tests}. The tests were built to evaluate the key elements of our concept. To this end, the tests mainly focus on:

\begin{itemize}
    \item Flow test - determine whether the system can successfully mint a new NFTAA contract, emit a 'NewNFTAA' event, and check if the minted NFT is bonded.
    \item Validations - the system ensures that transactions attempting to mint an NFTAA contract with an empty or excessively large description are correctly reverted.
    \item Staking - ensures that users can stake their funds, increase their stakes, and unstake them. Additionally, it verifies that users cannot unstake before the system reaches the unlock time.
    \item Attributes - the system's ability to return the proper bound NFT ID and NFTAA note is tested in these tests.
    \item Proxy - examine if the system correctly revokes actions taken by parties other than the contract owner and whether it issues a "ProxyResponse" event whenever a legitimate function call performed via the proxy is received.
\end{itemize}

Eleven subtests passed the testing phase. By running these automated tests, we can ensure that the smart contract system performs as expected and adheres to our expectations.

\subsection{Requirements evaluation}
%\subsubsection{Evaluation of functional requirements}
The evaluation of requirements, as depicted in Table \ref{tab:funreqeval}, demonstrates the successful implementation of key features within the system. The initial requirement (Request 1.) ensuring users can connect their wallets has been met, establishing a fundamental connection between users and the platform. Subsequently, the creation of NFTAAs (Non-Fungible Token Asset Accounts) is confirmed successful (Request 2.), emphasizing the system's capability to generate unique and distinct token accounts. Transactions are seamlessly facilitated with the successful transfer of NFTAAs between wallets (Request 3.), affirming the platform's robustness in handling asset transfers. Stakeholder engagement is supported through functionalities such as staking (Request 4.) and unstaking (Request 5.) via NFTAA, reflecting the system's commitment to user participation in blockchain-based activities.
Moreover, users can conveniently monitor their NFTAAs (Request 6.), fostering transparency and awareness of individual token holdings. The system's ability to verify ownership of NFTAA (Request 7.) adds an additional layer of security and accountability. The rest of the requirements are non-functional. We did a thorough, smart contract security audit by tools Slither\footnote{https://github.com/crytic/slither} and MythX\footnote{https://mythx.io/} to evaluate non-functional requirement number fourteen. There were no critical or severe issues in the security model. As a result, the acceptance testing of the non-functional requirements is successful. The evaluation underscores the system's effective incorporation of essential functionalities, meeting or exceeding the specified requirements and providing users with a comprehensive and reliable blockchain experience.

\begin{table}[!ht]
\centering
\caption{Evaluation of requirements}
\resizebox{\columnwidth}{!}{%
\begin{tabular}{|c|c|c|}
\hline
\textbf{Request identifier} & \textbf{Description of the requirement}                                                                     & \textbf{Evaluation} \\ \hline
1. & The system shall allow users to connect their wallet & Successful   \\ \hline
2. & The system shall allow users to create an NFTAA      & Successful   \\ \hline
3.                          & \begin{tabular}[c]{@{}c@{}}The system shall allow users to transfer NFTAA \\ to another wallet\end{tabular} & Successful            \\ \hline
4. & The system shall allow users to do stake via NFTAA   & Successful   \\ \hline
5. & The system shall allow users to do unstake via NFTAA & Successful   \\ \hline
6. & The system shall allow users to check their NFTAAs   & Successful \\ \hline
7.                          & \begin{tabular}[c]{@{}c@{}}The system shall allow users to check \\ owner of some NFTAA\end{tabular}        & Successful          \\ \hline
8.                          & The system core functionality shall be implemented by smart contract & Successful            \\ \hline
9. & \begin{tabular}[c]{@{}c@{}}The system shall provide a user-friendly interface, with clear and \\ concise instructions and error messages\end{tabular} & Successful \\ \hline
10. & \begin{tabular}[c]{@{}c@{}}The system shall be scalable to handle a large number of users \\ and data storage requirements\end{tabular}               & Successful \\ \hline
11.                          & The system shall be used by any wallet                           & Successful            \\ \hline
12.                          & The system shall be compatible with modern web browsers              & Successful            \\ \hline
13. & \begin{tabular}[c]{@{}c@{}}The system shall be compliant with relevant industry standards \\ and best practices\end{tabular}                          & Successful \\ \hline
14.                          & The system shall be secure and protected against cyber-attacks       & Successful          \\ \hline
\end{tabular}%
}
\label{tab:funreqeval}
\end{table}

%\subsubsection{Evaluation of non-functional requirements}

% Table \ref{tab:notfunreqeval} provides a brief overview of the evaluation of non-functional requirements. 

% \begin{table}[h!]
% \centering
% \resizebox{\columnwidth}{!}{%
% \begin{tabular}{|c|c|c|}
% \hline
% \textbf{Request identifier} & \textbf{Description of the requirement}                              & \textbf{Evaluation} \\ \hline
% 1.                          & The system core functionality shall be implemented by smart contract & Successful            \\ \hline
% 2. & \begin{tabular}[c]{@{}c@{}}The system shall provide a user-friendly interface, with clear and \\ concise instructions and error messages\end{tabular} & Successful \\ \hline
% 3. & \begin{tabular}[c]{@{}c@{}}The system shall be scalable to handle a large number of users \\ and data storage requirements\end{tabular}               & Successful \\ \hline
% 4.                          & The system shall be used by any wallet                           & Successful            \\ \hline
% 5.                          & The system shall be compatible with modern web browsers              & Successful            \\ \hline
% 6. & \begin{tabular}[c]{@{}c@{}}The system shall be compliant with relevant industry standards \\ and best practices\end{tabular}                          & Successful \\ \hline
% 7.                          & The system shall be secure and protected against cyber-attacks       & Successful          \\ \hline
% \end{tabular}%
% }
% \caption{Evaluation of non-functional requirements}
% \label{tab:notfunreqeval}
% \end{table}

\section{Conclusion and future work} \label{sec:summary}
Non-fungible tokens as accounts have massive potential in the Ethereum ecosystem.
They can provide a valuable asset that can be used as an account in the runtime and traded on the market.
One of the best features of the NFTAA system is that any account type can cooperate with NFTAA: externally owned accounts, smart contracts, proxies, or multi-signature accounts.
Staking is an excellent use case example for NFTAA.
Unlike classic liquid staking, NFTAA does not rely on any DAO, multi-signature accounts, or entity that manages the token; instead, the owners are in full custody of their assets.
Additionally, the network does not lose any security since no funds are unstaked and can be withdrawn. 
Moreover, non-fungible tokens can be used in other DeFi applications, such as lending, borrowing, and derivatives, so they do not lose any flexibility.
From the implementation perspective, the NFTAA can be implemented with current and available Solidity tooling. However, the applicability of NFTAA is not only in staking use case but also company owning, asset and portfolio management, delegation of rights to another wallet, DAO governance, or even identity and role management (e.g., you could have a game profile or persona linked to the NFT and in future sell it with every item in the itinerary). The possibilities of NFTAA are nearly endless. However, the key focus area of this research was the area of staking. We are opening new business opportunities with the new data availability layers and their re-staking mechanism.

 Our solution architecture is designed to be used with layer two solutions that are an alternative way of expanding Ethereum scalability. Our assessment shows feasibility and scale in the current Ethereum ecosystem, with a commitment to address scale concerns more in-depth in future revisions.

In summary, non-fungible tokens as accounts, specifically in the context of proxy-staking or token-bound accounts, introduce a paradigm shift in account management and staking within blockchain ecosystems. Integrating unique NFTs with account functionalities enhances personalization, participation, and governance within decentralized networks. It reflects the evolving landscape of blockchain technology, where innovative applications continually redefine the boundaries of traditional concepts.
This paper offers design and architecture decisions with a simple implementation, which allows developers and researchers to choose the best solution for their needs.

Future work includes examining other use cases for NFTAA and, most importantly, conducting a detailed security assessment to identify and mitigate potential risks. We know that staking through NFTAA reduces asset loss and network security risks compared to externally owned accounts. By decoupling the stake from the owner, we make staking delegation more effective because of the fact that NFTAA serves as a proxy for externally owned accounts.

Moreover, we would like to observe the usage of NFTAA in different networks, e.g., the Polkadot ecosystem, where there is already an implementation of proxy accounts and the unstacking lasts 28 days. 
Implementing NFTAA in the Substrase environment means bringing this concept to a completely different technology. Also, the plan is to extend the research to more enterprise use cases with great potential for business and blockchain fusion.

\section*{Acknowledgment}
This work was supported by the Slovak Research and Development Agency under Contract no. APVV-20-0338.

\printbibliography

\end{document}